\documentclass[twocolumn]{aastex62}

\graphicspath{{./}{figures/}}

\received{02-15-2019}
\revised{03-05-2019}
\accepted{03-08-2019}
\submitjournal{ApJ}

\shorttitle{MESAS: Sirius A}
\shortauthors{White et al.}

\begin{document}

\title{The MESAS Project: Long wavelength follow-up observations of Sirius A}

\correspondingauthor{Jacob Aaron White}
\email{jacob.white@csfk.mta.hu}

\author[0000-0001-8445-0444]{Jacob Aaron White}
  \affiliation{Konkoly Observatory,
  Research Centre for Astronomy and Earth Sciences,
  Hungarian Academy of Sciences,
  Konkoly-Thege Mikl\'os \'ut 15-17, 1121 Budapest, Hungary}

\author{J. Aufdenberg}
\affiliation{Physical Sciences Department,
Embry-Riddle Aeronautical University,
600 S Clyde Morris Blvd.,
Daytona Beach, FL 32114, USA}

\author{A.~C.~Boley}
\affil{Department of Physics and Astronomy,
University of British Columbia,
6224 Agricultural Rd.,
Vancouver, BC V6T 1T7, Canada}

\author{M. Devlin } 
\affiliation{Department of Physics and Astronomy,
University of Pennsylvania,
209 South 33rd Street,
Philadelphia, PA 19104, USA}

\author{S. Dicker}
\affiliation{Department of Physics and Astronomy,
University of Pennsylvania,
209 South 33rd Street,
Philadelphia, PA 19104, USA}

\author{P. Hauschildt}
\affiliation{Hamburger Sternwarte,
 Gojenbergsweg 112,
 21029 Hamburg, Germany}

\author{A.~G.~Hughes}
\affil{Department of Physics and Astronomy,
University of British Columbia,
6224 Agricultural Rd.,
Vancouver, BC V6T 1T7, Canada}

\author{A. M. Hughes}
\affiliation{Department of Astronomy,
Van Vleck Observatory,
Wesleyan University,
96 Foss Hill Dr.,
Middletown, CT 06459, USA}

\author{B. Mason}
\affiliation{National Radio Astronomy Observatory,
520 Edgemont Road,
Charlottesville, VA 22903, USA}

\author{B. Matthews}
\affiliation{Herzberg Institute,
 National Research Council of Canada,
 5071 W. Saanich Road,
 Victoria, BC V9E 2E7, Canada}

\author{A. Mo\'or}
  \affiliation{Konkoly Observatory,
  Research Centre for Astronomy and Earth Sciences,
  Hungarian Academy of Sciences,
  Konkoly-Thege Mikl\'os \'ut 15-17, 1121 Budapest, Hungary}

\author{T. Mroczkowski} 
\affiliation{ESO-European Organization for Astronomical Research in the Southern hemisphere,
Karl-Schwarzschild-Str. 2,
D-85748 Garching b. Munchen, Germany}

\author{C. Romero} 
\affiliation{Department of Physics and Astronomy,
University of Pennsylvania,
209 South 33rd Street,
Philadelphia, PA 19104, USA}

\author{J. Sievers}
\affiliation{McGill University,
3600 University Street,
Montreal, QC H3A 2T8, Canada}

\author{S. Stanchfield}
\affiliation{Department of Physics and Astronomy,
University of Pennsylvania,
209 South 33rd Street,
Philadelphia, PA 19104, USA}

\author{F. Tapia}
\affiliation{Posgrado en Ciencias de la Tierra, 
Universidad Nacional Autónoma de M\'exico, 
Morelia, 58190, M\'exico}
\affiliation{Escuela Nacional de Estudios Superiores Unidad Morelia, 
Universidad Nacional Autónoma de M\'exico, 
Morelia, 58190, M\'exico}

\author{D. Wilner}
\affiliation{Center for Astrophysics | Harvard \& Smithsonian,
60 Garden Street,
Cambridge, MA 02138, USA}



\begin{abstract}

Modeling the submillimeter to centimeter emission of stars is challenging due to a lack of sensitive observations at these long wavelengths. We launched an ongoing campaign to obtain new observations entitled Measuring the Emission of Stellar Atmospheres at Submillimeter/millimeter wavelengths (MESAS). Here we present ALMA, GBT, and VLA observations of Sirius A, the closest main-sequence A-type star, that span from 1.4 to 9.0 millimeters. These observations complement our previous millimeter data on Sirius A and are entirely consistent with the PHOENIX stellar atmosphere models constructed to explain them. We note that accurate models of long wavelength emission from stars are essential not only to understand fundamental stellar processes, but also to determine the presence of dusty debris in spatially unresolved observations of circumstellar disks.

\end{abstract}


\section{Introduction} \label{sec:intro}

The submillimeter to centimeter (submm-cm) emission of stars is largely unconstrained for spectral types different from the Sun \citep[see, e.g., ][for other G-type stars]{liseau15}. Due to the presence of a corona, the Sun can have a brightness temperature (T$_{\rm B}$) that dips below the optical photosphere T$_{\rm B}$ in the far-infrared. It then becomes much hotter and difficult to model at mm-cm wavelengths \citep[e.g.,][]{loukitcheva, wang, wedemeyer, delaluz14}. A Solar spectrum may well represent that of a typical G-type star, but cannot be used for all spectral types due to differences from stellar magnetic fields, surface convection, and atmospheric structure. Nevertheless, due to the lack of observations of different spectral types, the Sun can still serve as an illustrative example of long wavelength stellar emission of similar stars. In order to test atmospheric structure models and their applicability to, e.g., search for excess emission from circumstellar debris, observations of a broad range of spectral types is required.

Asteroids and comets are the leftovers of the planet formation process. As these objects dynamically evolve, they can serve as a new source of $\mu$m to cm sized debris that would have otherwise been cleared out of the disk \citep[for a general overview of debris disks, see][]{hughes18}. The presence of this circumstellar debris is commonly inferred through a pronounced excess over the expected stellar emission. Debris disks can be difficult to spatially separate from the host star for all but the closest stellar neighbors. The common approach of extrapolating the brightness temperature observed at $\rm \sim \mu$m wavelengths to millimeter wavelengths may under or over predict the flux contribution of the star. As debris disks are most commonly found around A-type stars \citep{su06,thureau}, assuming a Solar like spectrum will typically not be accurate either. Therefore a poor characterization of the host star's submm-cm flux can non-trivially affect measurements of the occurrence and abundance of circumstellar debris.

Sirius A is a 225 - 250 Myr A1Vm star \citep{adelman, liebert, bond17} that has no observed circumstellar debris. Located only $2.64\pm0.01$ pc \citep{vanleeuwen} from our Solar System, this bright star is an excellent target to observe and model the submm-cm stellar emission of A-type stars. \citet{white18} presented long wavelength observations of Sirius A with the \textit{James Clerk Maxwell Telescope} (JCMT), \textit{Submillimeter Array} (SMA), and the \textit{NSF's Karl G. Jansky Very Large Array} (VLA) and provided PHOENIX models \citep{hauschildt} of its stellar photosphere. The models showed that an extrapolation of short wavelength data could not accurately describe the submm-cm emission of Sirius A, which has implications for similar types of debris-hosting stars such as Fomalhaut \citep{white_dis}. 

In this paper, we present follow-up observations of Sirius A. These data are part of an ongoing effort to characterize stellar atmospheres at submm-cm wavelengths through the MESAS Project (Measuring the Emission of Stellar Atmospheres at Submillimeter/centimeter wavelengths). 

\section{Observations}\label{sec:obs}

Previous long wavelength observations of Sirius A are summarized in \citet{white18} and are shown in Fig.\,\ref{tb_plot} (black circles). The new observations are detailed below.

\subsection{ALMA} \label{sec:alma}

We observed Sirius A with the \textit{Atacama Large Millimeter Array} (ALMA) during Cycle 5 (ID 2017.1.00698.S, PI White). The observations in ALMA Bands 3, 4, and 5 were taken in 3 separate executions blocks (EBs) from January through August. The observations were centered on Sirius A using J2000 coordinates RA = 06 h 45 m 08.30 s and $\delta = -16^{\circ} ~43' ~18.91''$. Each band utilized an instrument configuration with a total bandwidth of 8 GHz split among 4 spectral windows (SPW). Each SPW has $128\times 15.625$ MHz channels for a total bandwidth of 2 GHz.  All of the data were reduced using the Common Astronomy Software Applications ({\scriptsize CASA 5.4.0}) pipeline \citep{casa_reference}, which included water vapor radiometer (WVR) calibration; system temperature corrections; flux and bandpass calibration; and phase calibration.

The Band 5 observations were made on 2018 August 22 for 20.24 min (5.1 min on-source). This EB used a 43 antenna configuration with baselines ranging from 15 m to 457 m. The SPWs were centered at 196 GHz, 198 GHz, 208 GHz, and 210 GHz, giving an effective continuum frequency of 203 GHz (1.48 mm). The flux and bandpass were calibrated with quasar J0750+1231, the phase with J0648-1744, and the WVR with J0647-1605. The average precipitable water vapor (PWV) was 0.67 mm during the observations. These Band 5 data achieve a RMS sensitivity of $0.040~\rm mJy~beam^{-1}$ in the CLEANed image. The size of the resulting synthesized beam is $1.46''\times0.85''$ at a position angle of $-75.4^{\circ}$, corresponding to $\sim3.1$ au at the system distance of 2.64 pc.

The Band 4 observations were made on 2018 April 21 for 21.0 min (5.1 min on-source). This EB used a 43 antenna configuration with baselines ranging from 15 m to 500 m. The SPWs were centered at 143 GHz, 145 GHz, 155 GHz, and 157 GHz, giving an effective continuum frequency of 150 GHz (2.0 mm). The flux and bandpass were calibrated with quasar J0522-3627, the phase with J0654-1053, and the WVR with J0648-1744. The average PWV was 2.58 mm during the observations. These Band 4 data achieve a RMS sensitivity of $0.029~\rm mJy~beam^{-1}$ in the CLEANed image. The size of the resulting synthesized beam is $1.43''\times1.17''$ at a position angle of $-89.5^{\circ}$, corresponding to $\sim3.4$ au.

The Band 3 observations were made on 2018 January 25 for 19.75 min (5.12 min on-source). This EB used a 43 antenna configuration with baselines ranging from 15 m to 1397 m. The SPWs were centered at 93 GHz, 95 GHz, 105 GHz, and 107 GHz, giving an effective continuum frequency of 100 GHz (3.0 mm). The flux and bandpass were calibrated with quasar J0522-3627, the phase with J0654-1053, and the WVR with J0653-0625. The average PWV was 5.51 mm during the observations. These Band 3 data achieve a RMS sensitivity of $0.032~\rm mJy~beam^{-1}$ in the CLEANed image. The size of the resulting synthesized beam is $1.02''\times0.69''$ at a position angle of $68.2^{\circ}$, corresponding to $\sim2.3$ au.

\subsection{VLA} \label{sec:vla}

The data from the VLA were acquired in Semester 18A on 2018 September 07 and 2018 September 11 (ID 18A-328, PI White). The observations were centered on Sirius A using J2000 coordinates RA = 06 hr 45 min 08.30 sec and $\delta = -16^{\circ} ~43' ~18.91''$. The observations were taken in the D antenna configuration with 26 and 25 antennas, respectively, and had baselines ranging from 0.035 to 1.03 km. 

The Scheduling Block (SB) for each observation was setup identically to the previous VLA observations of Sirius A \citep{white18} to observe 6.7 and 9.0 mm (45 and 33 GHz) as close to simultaneously as possible. The 33 GHz data used the Ka Band tuning setup with $4\times2.048$ GHz basebands and rest frequency centers of 28.976, 31.024, 34.976, and 37.024 GHz. This gives an effective frequency of 33 GHz (9.0 mm) for the Ka band. The 45 GHz data used the Q Band tuning setup with $4\times2.048$ GHz basebands and rest frequency centers of 41.024, 43.072, 46.968, and 48.976 GHz. This gives an effective frequency of 45 GHz (6.7 mm) for the Q band. Quasar J0650-1637 was used for bandpass and gain calibration. Quasar 3C48 was used as a flux calibration source. Data were reduced using the  {\scriptsize CASA 5.4.0} pipeline, which included bandpass, flux, and phase calibrations. 

The absolute flux calibration of the VLA at these wavelength is typically $\sim5\%$. However, the flux calibrator, 3C48, was undergoing a flare at the time of the observations of Sirius A. This leads to a larger uncertainty\footnote{For a note on the VLA flux calibration uncertainty, see science.nrao.edu/facilities/vla/docs/manuals/oss/performance/fdscale. } in the flux calibration solution and, after consultation with the HelpDesk, we adopt a flux calibration uncertainty of $20\%$ for the VLA observations.

The data were imaged with a natural weighting and cleaned using {\scriptsize CASA}'s \textit{CLEAN} algorithm down to a threshold of 1/2 the RMS noise. The 2018 September 07 observations achieve a sensitivity of $0.022~\rm mJy~beam^{-1}$ and $0.015~\rm mJy~beam^{-1}$ for the 6.7 and 9.0 mm data, respectively. The size of the resulting synthesized beam is $3.24''\times1.16''$ ($\sim5.8$ au) at a position angle of $-17.5^{\circ}$ and $3.80''\times1.44''$ ($\sim6.9$ au) at a position angle of $-14.5^{\circ}$, respectively. The 2018 September 11 observations achieve sensitivities of $0.028~\rm mJy~beam^{-1}$ and $0.014~\rm mJy~beam^{-1}$, respectively. The size of the resulting synthesized beam is $3.04''\times1.39''$ ($\sim5.9$ au) at a position angle of $-10.6^{\circ}$ and $3.81''\times1.99''$ ($\sim7.7$ au) at a position angle of $3.0^{\circ}$, respectively.

\subsection{GBT} \label{sec:gbt}

Sirius A was observed by MUSTANG2 on the \textit{Green Bank Telescope} (GBT) on the nights of the 2018 February 01 and 2018 December 10. The MUSTANG2 instrument has a 75 GHz to 105 GHz bandpass \citep{dicker14}. The total on-source time was 104 min split roughly equally between nights.  A $2.5'$ radius daisy scan pattern lasting 480 s was used.  Every 3 scans, a secondary calibrator close to Sirius A (J0609-1542 in February and J0607-0834 in December) was observed with a similar but shorter scan.  The secondary calibrator observations were used to track pointing offsets, check the focus, and obtain an absolute calibration.  These secondary calibrator data were reduced using the same data pipeline and a scaling factor was applied for amplitude corrections.  The same scaling factor, extrapolated between scans, was used on the Sirius A data.

At the beginning and end of the night a collection of ALMA flux calibrators (J0423-0120, J1058+0133, J0750+1231 and J0510+1800) and the planet Uranus were also observed.  At least 3 independent back-to-back observations of one of these sources and our secondary calibrators were made each night.  Opacity corrections were made using $\tau$ as estimated from atmospheric models and archival weather data.  Taking into account the variability of the opacity, errors in the Gaussian fits to the sources, errors in measurements, errors in the ALMA fluxes, and errors resulting in the extrapolation of these fluxes into the MUSTANG2 band (see Sec.\,\ref{sec:model}) we measured J0609-1542 to be $1.54\pm0.1$ Jy (February 02) and J0607-0834 to be $2.15\pm0.06$ Jy (December 10).

\section{Visibility Model Fitting}\label{sec:model}

Sirius A is effectively a point source ($0.00602''$) when considering the size ALMA and the VLA's synthesized beams and the resolution of GBT. Therefore, in order to accurately model the flux from the interferometric observations, we  model Sirius A as a point source. As in \citet{white18}, we use the {\scriptsize CASA} task \textit{uvmodelfit} to obtain a flux at each wavelength. This approach fits a point source to the visibilities of a given data set. A minimum $\chi^{2}$ is converged on through an iterative procedure. The ALMA observations achieved a high signal-to-noise ratio in each band ($>50$) allowing for an accurate model fitting in each of the 4 SPWs in Bands 3-5. The results of the model fitting are summarized in Table\,\ref{fit_par}.

To model Sirius A's flux from the GBT observations, we generated a map of all the GBT data and fit a Gaussian at the location of Sirius A. This yields a total flux of $0.90\pm0.11$ mJy. The uncertainty is derived from the fit and calibrations uncertainties added in quadrature.  For confirmation, fake sources of known amplitude were added to the raw timestreams and were recovered with high accuracy.  Independent fits of each of the two nights data gave results consistent with no source variability to within the uncertainties.

The flux uncertainties in Table\,\ref{fit_par} include the absolute flux calibration the $\sigma_{RMS}$ added in quadrature. For ALMA, we adopt an absolute flux calibration uncertainty of $10\%$ and for VLA $20\%$ (see Sec.\,\ref{sec:vla}).

\begin{table*}

\centering 
\begin{tabular}{c  c  c  c  c  c  c} 
\hline\hline \ 
   	Wavelength  & Facility & Date & Flux Calibrator & Flux  & Uncertainty & Reduced $\chi^{2}$\\
   		$[$mm] &  & YYYY MMM DD & & [$\rm mJy$] &  [$\rm mJy$] & \\
   	\hline 
 
	1.43 & ALMA  & 2018 AUG 22  & J0750+1231 & 5.60  & 0.60  & 1.34\\
	1.44 & ALMA  & 2018 AUG 22  & J0750+1231 & 5.57  & 0.60  & 1.35\\
	1.51 & ALMA  & 2018 AUG 22  & J0750+1231 & 4.95  & 0.50  & 1.33\\
	1.53 & ALMA  & 2018 AUG 22  & J0750+1231 & 4.87  & 0.50  & 1.34 \\
	1.91 & ALMA   & 2018 APR 21  & J0522-3627 & 3.32  & 0.33  & 3.57\\
	1.93 & ALMA   & 2018 APR 21  & J0522-3627 & 3.22  & 0.32  & 3.85\\
	2.07 & ALMA   & 2018 APR 21  & J0522-3627 & 2.83  & 0.29  & 3.98\\
	2.10 & ALMA   & 2018 APR 21  & J0522-3627 & 2.73  & 0.27  & 3.58\\
	2.80 & ALMA   & 2018 JAN 25  & J0522-3627 & 1.73  & 0.18  &3.75\\
	2.86 & ALMA   & 2018 JAN 25  & J0522-3627 & 1.41  & 0.15  & 3.83\\
	3.16 & ALMA   & 2018 JAN 25  & J0522-3627 & 1.32  & 0.14  &3.78\\
	3.22 & ALMA   & 2018 JAN 25  & J0522-3627 & 1.35  & 0.14  & 3.73\\
	3.33 & GBT & 2018 FEB 01, DEC 10  & J0423-0120$^{a}$ & 0.90 & 0.11  & -\\
	6.7 & VLA   & 2018 SEP 07  & 3C48$^{b}$ & 0.212 & 0.047 & 3.42\\
	6.7 & VLA   & 2018 SEP 11  & 3C48$^{b}$ & 0.193 & 0.049 & 3.41\\
	9.0 & VLA   & 2018 SEP 07  & 3C48$^{b}$ & 0.125 & 0.029 & 4.46\\
	9.0 & VLA   & 2018 SEP 11  & 3C48$^{b}$ & 0.127 & 0.030 & 4.25\\
	
\hline	
    
\end{tabular}
\caption{Summary of observations and best fit flux values of Sirius A. The GBT flux value is from fitting a Gaussian to the map. The ALMA and VLA flux values were calculated using {\scriptsize CASA}'s \textit{uvmodelfit}. The flux uncertainties are the $\sigma_{\rm RMS}$ of the images and the absolute flux calibration uncertainties added in quadrature. For ALMA, a 10\% flux calibration uncertainty is used and for the VLA a 20\% flux calibration uncertainty is used for the VLA data.  The uncertainties from \textit{uvmodelfit} are not quoted as they can be underestimated up to a factor of $\sqrt{\chi^{2}_{\rm red}}$. (a) Multiple flux calibrators were used to calibrate the GBT data (see Sec.\,\ref{sec:gbt}). (b) The quasar 3C48, which was used to calibrate the flux from the VLA observations, has been undergoing a flare since early 2018 and therefore contribute to a larger absolute flux calibration uncertainty than is typical for these wavelength data (see Sec.\ref{sec:vla}).
}
\label{fit_par}
\end{table*}

\section{Discussion}\label{sec:disc}

The data presented here, in conjunction with the results from \citet{white18}, provide the first reliable submm-cm flux estimates of an A-type main-sequence star.

Sirius A's white dwarf companion, Sirius B, was not detected in any of the observations. As it has no detectable IR excess \citep{skemer11}, the expected flux of Sirius B can be calculated using the effective temperature of T$_{eff}=25369~\rm K$ and radius of $\rm 0.008098~R_{\odot}$ \citep{bond17}. The flux is less than the $\sigma_{RMS}$ for all our observations except for the $\rm1.48~mm$ ALMA observations, where the expected emission is roughly the same as the noise. Sirius B was not present in any of the deconvolved images. In addition, the orbital parameters of Sirius B \citep{bond17} and the $\sim50~\rm yr$ period have it nearing aphelion with a radial separation of $\sim6''$. While this separation will be within the FOV of all the facilities used, the resolution of all the interferometic observations is sufficient to spatially separate the emission from the two stars. The resolution of the GBT is not adequate to separate the two stars, but the expected emission of Sirius B is much lower than the sensitivity of the observations. 

\begin{figure*}
\centering
\includegraphics[width=0.75\textwidth]{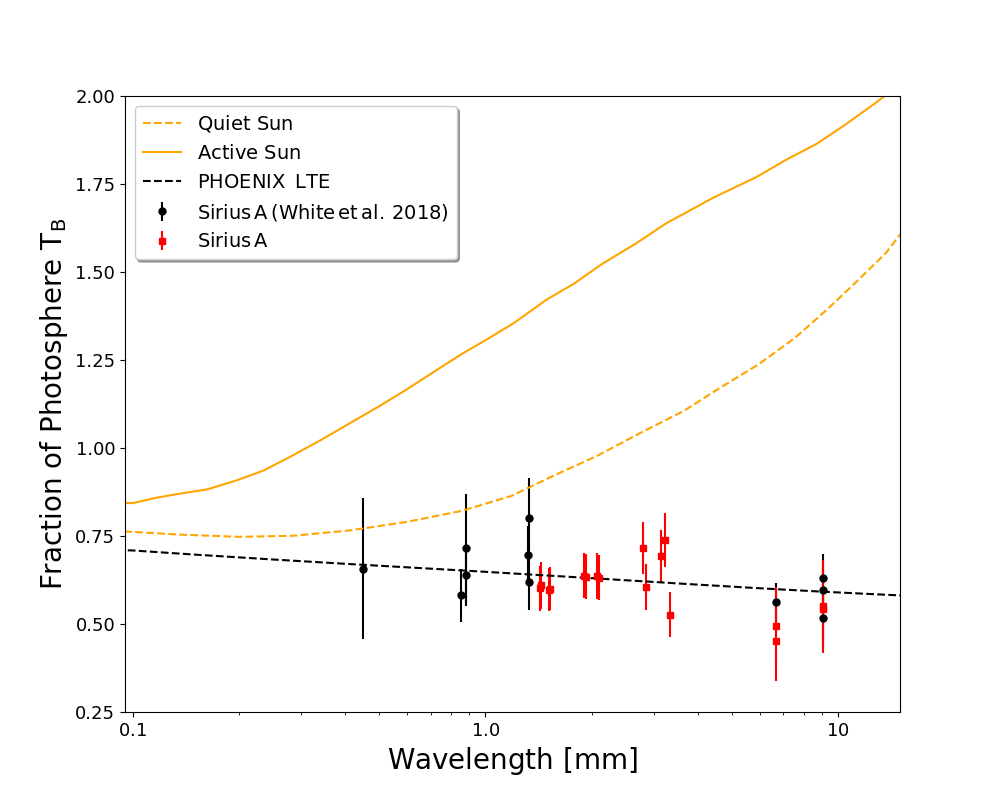} 
\caption{Submm-cm observations of Sirius A. The two orange curves represent models of the Sun at maximum activity (solid line) and minimum activity (dashed line) from \citet{loukitcheva}. A Solar-like spectrum is not expected for A-type stars and is only included for illustrative purposes. The observations of Sirius A from \citet{white18} are denoted as black circles. The newly presented ALMA, GBT, and VLA observations of Sirius A are denoted as red squares. The two black curves are PHOENIX models of Sirius A's atmosphere with a LTE model (dashed black line) as presented in \citet{white18}.
\label{tb_plot}
}
\end{figure*}

In Fig.\,\ref{tb_plot}, we include the LTE atmosphere model of Sirius A from \citet{white18}. This model generates a synthetic spectra of Sirius A's photosphere from a spherical 1D PHOENIX code \citep[version 17.1; see][for an earlier code version]{hauschildt10}. This modeling procedure adopted a similar approach as \citet{husser13}, and updated it to converge 192 atom/ion species with 37,419 levels in statistical equilibrium and 1,283,018 total spectral lines \citep[see][for a detailed overview of the LTE and non-LTE PHOENIX models of Sirius A]{white18}. 

Our first observations of Sirius A were with the SMA in January 2017 \citep{white18} and our most recent were with GBT in December 2018. All of the data over this $\sim$2 yr period are largely consistent with the PHOENIX LTE atmosphere model (Fig.\,\ref{tb_plot}). This stability in the emission indicates that there is likely no variability in Sirius A's long wavelength emission on at least a two year cycle at the level of the flux calibration accuracy (i.e., the $20\%$ level). Indeed, we can not exclude the possibility of short-lived mm flares \citep[e.g.,][]{furuya03, salter10, macgregor18}. However, such stellar activity is not expected for main sequence A-type stars. As variability has been seen at long wavelengths for Herbig Ae/Be type stars \citep[e.g., at 9.0 mm in HD141569;][]{white_141569}, long term monitoring of Sirius A's emission is required to conclusively determine its activity cycle, should one exist.

The submm-cm spectrum of Sirius A can be used as a template for other A-type stars. To approximate the flux of Sirius A within the wavelength range of our JCMT to VLA observations (i.e., $0.45 - 9.0$ mm) we fit a power law to the observed brightness temperature spectrum of the form:
\begin{equation}
    T_{obs} = C_{0} T_{phot} \Big(\frac{\lambda}{0.45\,\rm mm}\Big)^\alpha,
\end{equation}
where C$_{0}$ is a constant, $T_{phot}$ is the optical photosphere temperature, $\lambda$ is the wavelength in mm, and $\alpha$ is the Temperature spectral index. Using a non-linear least squares model fitting procedure with the Python {\scriptsize SciPy} function \textit{curve\_fit}, we find best fit values of C$_{0} = 0.67\pm 0.03$ and $\alpha = 0.05\pm 0.02$. This approach has the benefit of predicting the stellar flux for any A-type star similar to Sirius A based only on a few parameters: the optical photosphere temperature (T$_{phot}$), stellar radius (R$_{*}$), and distance (d). The submm-cm flux, $F_{*}$, can then be estimated by assuming the stellar emission is in the Rayleigh-Jeans limit and substituting $T_{obs}$ for T$_{B}$:
\begin{equation}
F_{*} = A_{0} \Big(\frac{R_{*}}{d}\Big)^2 T_{phot} \lambda^{-2.052} \rm \,\,\, Jy,
\end{equation}
where A$_{0} \sim 5.6\times10^{9}$ is a collection of all the numerical coefficients and constants, and $\lambda$ is the wavelength in mm. In the absence of observational data, the $0.45 - 9.0$ mm flux of an early main sequence A-type star can be approximated by Eqn.\,2.

When studying unresolved circumstellar debris, a model of the stellar emission is required to accurately assess the amount of stellar excess. Since debris disks are most commonly found around A-type stars, Eqn.\,2 can be used in determining how much excess emission is present in the submm-cm. For example, Fomalhaut is a main sequence A3V star and is very similar to Sirius A.  In addition to Fomalhaut's well resolved outer debris ring, IR observations of the system are consistent with a secondary inner asteroid belt \citep[e.g.,][]{acke}. ALMA observations of the system at 0.87 and 1.3 mm did not, however, detect the central debris disk and observed a flux of 1.79 mJy and 0.88 mJy, respectively \citep{su16, white_fom}. Plugging in a stellar radius of $\rm 1.84~R_{\odot}$ \citep{mamajek}, a distance of 7.7 pc \citep{vanleeuwen}, and a photosphere temperature of 8650 K \citep{acke} to Eqn.\,2 yields a predicted flux of 1.88 mJy and 0.84 mJy for the 2 ALMA wavelengths, respectively. The absolute flux calibration uncertainty of ALMA at these wavelengths is $\sim10\%$, meaning that the observed flux is consistent with stellar emission alone. \citet{white_dis} find through SED modeling that the presence of stellar excess at IR wavelengths and lack of excess at $\sim$mm wavelengths could be consistent with an inner disk that is only populated by small grains that have migrated inwards from the outer debris belt through, e.g., Poynting-Robertson drag. 

While the spectral profile of Sirius A may be a good match for debris systems such as Fomalhaut, Sirius A can not necessarily be used as a template for all A-type stars, as a given star could deviate significantly from Sirius A's profile. Altair is a main sequence A star, but is a rapid rotator with a period of $\sim9$ hours \citep{peterson06, monnier07}. This rotation could potentially generate magnetic activity around the stellar equator, and indeed NOEMA observations from $0.87-3.0$ mm have indicated Altair has a remarkably different and more Solar-like T$_{B}$ spectrum (White et al. in prep). 

Therefore, a full catalog of stellar spectra that extend to cm wavelengths is necessary to accurately study unresolved debris features. Currently, the available data of debris-poor stars is largely limited by the long integration times required to obtain a useful SNR. Proposed future facilities such as the \textit{next generation Very Large Array} (ngVLA) will be a valuable to tool in extending the stellar models to $>$cm wavelengths \citep{white_ngvla}.

\section{Summary}\label{sec:sum}

In this paper, we presented new ALMA and GBT observations of Sirius A at $1.43-3.33$ mm and follow-up VLA observations at 6.7 mm and 9.0 mm. The newly acquired data are in good agreement with previous observations and PHOENIX models of the Sirius A's stellar atmosphere. Since these observations show no significant variability in the millimeter to centimeter emission from Sirius A over the two years that our observations sample, we can conclude that the emission is likely constant (within the 20\% calibration uncertainties). These data can be used as a long wavelength template of stellar emission for A-type stars, which is necessary to accurately study unresolved circumstellar material such as exozodiacal dust and debris disks. These results are part of an ongoing observational campaign entitled MESAS.

\acknowledgments

We thank the anonymous referee for the feedback on this manuscript. JAW acknowledges support from the European Research Council (ERC) under the European Union's Horizon 2020 research and innovation program under grant agreement No 716155 (SACCRED). AM acknowledges sopport from the Hungarian National Research, Development and Innovation Office NKFIH Grant KH-130526. FT acknowledges Conacyt-254497 (Ciencia B\'asica) and the Center of Supercomputing of the National Laboratory of Space Weather in M\'exico.

This paper makes use of the following ALMA data: ADS/JAO.ALMA\#2017.1.00698.S. ALMA is a partnership of ESO (representing its member states), NSF (USA) and NINS (Japan), together with NRC (Canada), MOST and ASIAA (Taiwan), and KASI (Republic of Korea), in cooperation with the Republic of Chile. The Joint ALMA Observatory is operated by ESO, AUI/NRAO and NAOJ. The National Radio Astronomy Observatory is a facility of the National Science Foundation operated under cooperative agreement by Associated Universities, Inc.

The Green Bank Observatory is a facility of the National Science Foundation operated under cooperative agreement by Associated Universities, Inc.  Support for the MUSTANG2 instrument team is provided by NSF/AST-1615604. 

The PHOENIX model calculations were performed at the RRZ of the Universit\"at Hamburg, H\"ochstleistungs Rechenzentrum Nord (HLRN), and  National Energy Research Supercomputer Center (NERSC), which is supported by the U.S. Office of Science of the DOE (No. DE-AC03-76SF00098). We thank all these institutions for a generous allocation of computer time. PHH acknowledges support from NVIDIA Corporation with the donation of a Quadro P6000 GPU used in this research.

\vspace{5mm}
\facilities{ALMA, GBT, VLA}


\software{{\scriptsize CASA 5.4.0} \citep{casa_reference} 
          }



\end{document}